\documentclass{webofc}
\usepackage[varg]{txfonts}   
\usepackage{hyperref}
\begin{document}
\title{\raggedright Parallelizing Air Shower Simulation for Background Characterization in IceCube}

\author{
  \firstname{Kevin} \lastname{Meagher}\inst{1}\fnsep\thanks{\email{kjmeagher@wisc.edu}}
  \and
  \firstname{Jakob} \lastname{van Santen}\inst{2}
  \,
  for the IceCube Collaboration\thanks{\protect\url{http://icecube.wisc.edu}}
}

\institute{
  \raggedright Dept. of Physics and Wisconsin IceCube Particle Astrophysics Center, University of Wisconsin,
  \linebreak Madison, WI 53706, USA.
  \and
  Deutsches Elektronen-Synchrotron DESY, Platanenallee 6, 15738 Zeuthen, Germany.
}

\abstract{%
The IceCube Neutrino Observatory is a cubic kilometer neutrino telescope 
located at the Geographic South Pole. For every observed neutrino event, 
there are over $10^{6}$ background events caused by cosmic ray air shower 
muons. In order to properly separate signal from background, it is 
necessary to produce Monte Carlo simulations of these air showers. 
Although to-date, IceCube has produced large quantities of background 
simulation, these studies still remain statistics limited. 
The first stage of simulation requires heavy CPU usage while 
the second stage requires heavy GPU usage. 
Processing both of these stages on the same node will result in an 
underutilized GPU but using different nodes will encounter bandwidth 
bottlenecks. Furthermore, due to the power-law energy spectrum of cosmic 
rays, the memory footprint of the detector response often exceeded the 
limit in unpredictable ways. This proceeding presents new 
client--server code which parallelizes the first stage onto multiple CPUs 
on the same node and then passes it on to the GPU for photon 
propagation. This results in GPU utilization of greater than 90\% as well 
as more predictable memory usage and an overall factor of 20 improvement 
in speed over previous techniques.
}
\maketitle
\section{Introduction}
\label{intro}

The IceCube Neutrino Observatory is a cubic kilometer telescope designed to observe neutrinos 
with energies of 100 GeV and above.
It was built by assembling 5160 Digital Optical Modules (DOMs) and placing them in the 
glacial ice of Antarctica between depths of 1450\,m and 2450\,m.
Each DOM contains a 25 cm photomultiplier tube (PMT), high voltage power supply, and digitization and communication electronics. 
The optical properties of the ice vary significantly over the range of the detector's depth.
These depth dependent optical properties make it so that simple models of light propagation in the ice are not accurate.
In order to obtain accurate simulation it is necessary to use Monte Carlo simulation to 
track individual photons which is computationally expensive.

The large size of the IceCube detector means that it also observes a high rate of muons from 
cosmic ray air showers.
For most analyses these muon events are considered undesirable background.
IceCube's trigger rate is approximately 3\,kHz, which is almost all muon events,
whereas atmospheric neutrinos are only detected at a rate of about 2\,mHz;
thus, any neutrino analysis requires a data reduction of at least 6 orders of magnitude.
In order to properly perform an analysis with this level of data reduction,
it is necessary to produce a large quantity of background simulation.
In Section \ref{old-simulation}, we will describe the former state of IceCube's cosmic ray simulation
and then in Section \ref{old-is-bad} we describe why this method uses computing resources inefficiently.
In Section \ref{client-server}, we describe a new client--server simulation stack and how it 
was designed to address the deficiencies of the old chain.
In Section \ref{results}, we show that the new chain shows significant improvement in efficiency
while yielding the same science results.
Finally, we discuss further possible improvements in Section \ref{future}.

\begin{figure}[t]
  \centering     
  \includegraphics[width=\textwidth,clip]{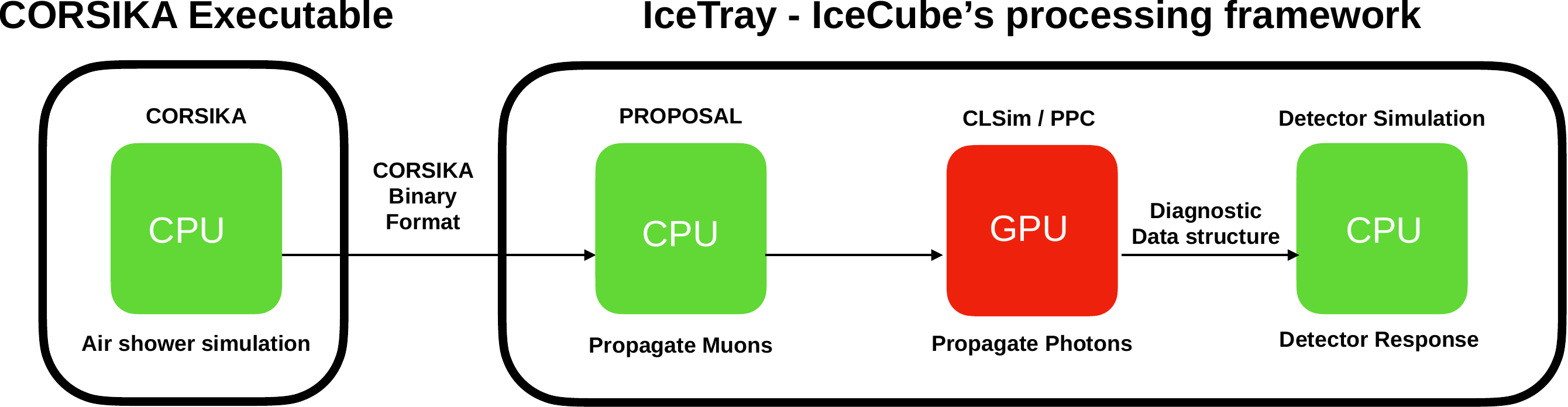}
  \caption{
    Simplified diagram of the old IceCube simulation chain.
    The direction and primary energy are sampled by \textsc{Corsika} before it simulates the air shower.
    The events are written to disk in \textsc{Corsika}'s binary format.
    The files are then read by \textsc{IceTray}~\cite{IceTray}, IceCube's processing framework.
    The muons from the air shower are then propagated in the ice by \textsc{Proposal}.
    The light produced by \textsc{Proposal} is then handed off the GPU to propagate photons.
    Finally the detector response is simulated.
    All stages can only perform work on one event at a time causing inefficiencies in the simulation.
    }
  \label{block-old}
\end{figure}

\section{Old Simulation Chain}
\label{old-simulation}

As shown in Figure \ref{block-old}, IceCube's simulation chain can be broken down into four components:
air shower propagation, muon propagation, photon propagation, and detector simulation.
Each stage is implemented using one or more modules in \textsc{IceTray}~\cite{IceTray},
IceCube's framework for serially processing data and simulation.
The air shower simulation starts with a cosmic ray primary and tracks all secondary particles
as they travel through the atmosphere where they interact with air molecules or decay.
This simulation is performed by \textsc{Corsika~7}~\cite{CORSIKA7}, a standard software package used throughout the cosmic ray physics
and high-energy astronomy community.
Muons and neutrinos are the only secondary particles that can reach the detector depth of IceCube.
Muons produced in the air shower are then tracked through the ice with \textsc{Proposal}~\cite{PROPOSAL3},
which was designed for efficiently propagating muons on length scales needed for deep ice detectors.
\textsc{Proposal} tracks muons by dividing their energy losses into continuous and stochastic losses;
the energy threshold between these two losses can be set based on the specific application:
A lower threshold for more accuracy and a higher threshold for faster compute times.
To save computing resources, the threshold is set to a higher energy for tracking from the surface to the detector.
Once the muon enters the detector volume, the threshold is lowered to get an accurate account
of energy losses.
Electromagnetic and hadronic secondaries that are generated by muons are tracked by 
code which uses parameterized light yields based on \textsc{Geant4} studies~\cite{CMC}.

These losses in the detector volume are then used as light sources for the photon propagator.
The optical properties of the ice are divided into vertical layers of 10\,m thickness having 
different coefficients of scattering and absorption. 
Light is propagated by using Monte Carlo techniques to calculate the distance to the next scatter 
for each photon and then picking a new angle based on the physics of scattering.
This process is then repeated until the photon intersects a detector or is absorbed.
When a photon intersects a detector, it creates a photoelectron with a probability
calculated from its position, direction, time, and wavelength.
This process is highly computationally intensive but also embarrassingly parallel
and thus benefits significantly from GPUs.
IceCube has two photon propagators \textsc{PPC}~\cite{PPC1,PPC2,PPCrepo} and \textsc{CLSim}~\cite{IceModel,CLSim} 
which can be used interchangeably. 

The final step is detector simulation, in which photoelectrons registered by the photon propagator
are converted into digital readings by the DOM hardware using parameterizations 
of the PMT response and front end electronics~\cite{IceCubePMT}.
Noise in the PMT and data acquisition~\cite{IceCubeDAQ} are also simulated. 
After this, simulation is then run through the exact same reconstructions and filtering
algorithms as experimental data.

\section{Problems with Old Simulation Chain}
\label{old-is-bad}

The simulation chain described above is effective at producing a simulation that is in agreement with data.
However, there are several inefficiencies that significantly slow down production.
\textsc{Corsika} and \textsc{Proposal} are highly stochastic and are not easily parallelizable, and so need to be run on CPUs.
Running a simulation chain with the GPU based photon propagation sandwiched between CPU intensive
code creates a number of challenges for large scale production of simulation.

IceCube uses a distributed processing system, called \textsc{IceProd}~\cite{iceprod},
which routinely runs jobs with dependency graphs, such that related jobs can run on different nodes in succession.
This makes it possible to run a CPU job and then send the results to a GPU job.
This has worked reasonably well historically, but has several problems in practice:

Since Muon propagation through the ice is highly stochastic, most muons do not reach the detector volume.
This creates an Input/Output bottleneck for low energy primaries as all the muons produced in the air shower
must be saved and transferred
from the first stage node to the second stage which can take up a significant portion of the processing 
time. 

Unfortunately, attempting to run the CPU stage and GPU stage on the same node
causes a different issue.
The time needed for CPU calculation exceeds the time for GPU, which means 
that the GPU will sit idle for a significant fraction of the time.
Another issue that arises is that in order to properly exploit the parallel nature of the GPU,
photons must be sent to the GPU as a fairly large batch,
and many events will have too few light sources to properly utilize the GPU.

In addition, the chain requires that all the information about an event to be stored in memory:
for high energy events, this can often exceed the available memory, typically 2GB. 
Jobs that exceed this limit will be halted by the computing orchestrator and must be resubmitted with a higher 
memory limit, resulting in wasted computing resources.
Due to the extremely stochastic nature of air showers and muon propagation and the power-law
primary spectrum, it is very hard to predict which events will exceed the memory limit.
This problem will only get worse as new computing resources come online.
Although the amount of memory in new computing infrastructure continues to rise,
the number of CPUs per node is increasing faster than memory.
The net result is that the amount of memory per processor core is expected to decrease over time.
A solution to this issue must be found in order to efficiently utilize future computing resources.

\begin{figure}[t]
  \centering     
  \includegraphics[width=0.68\textwidth,clip]{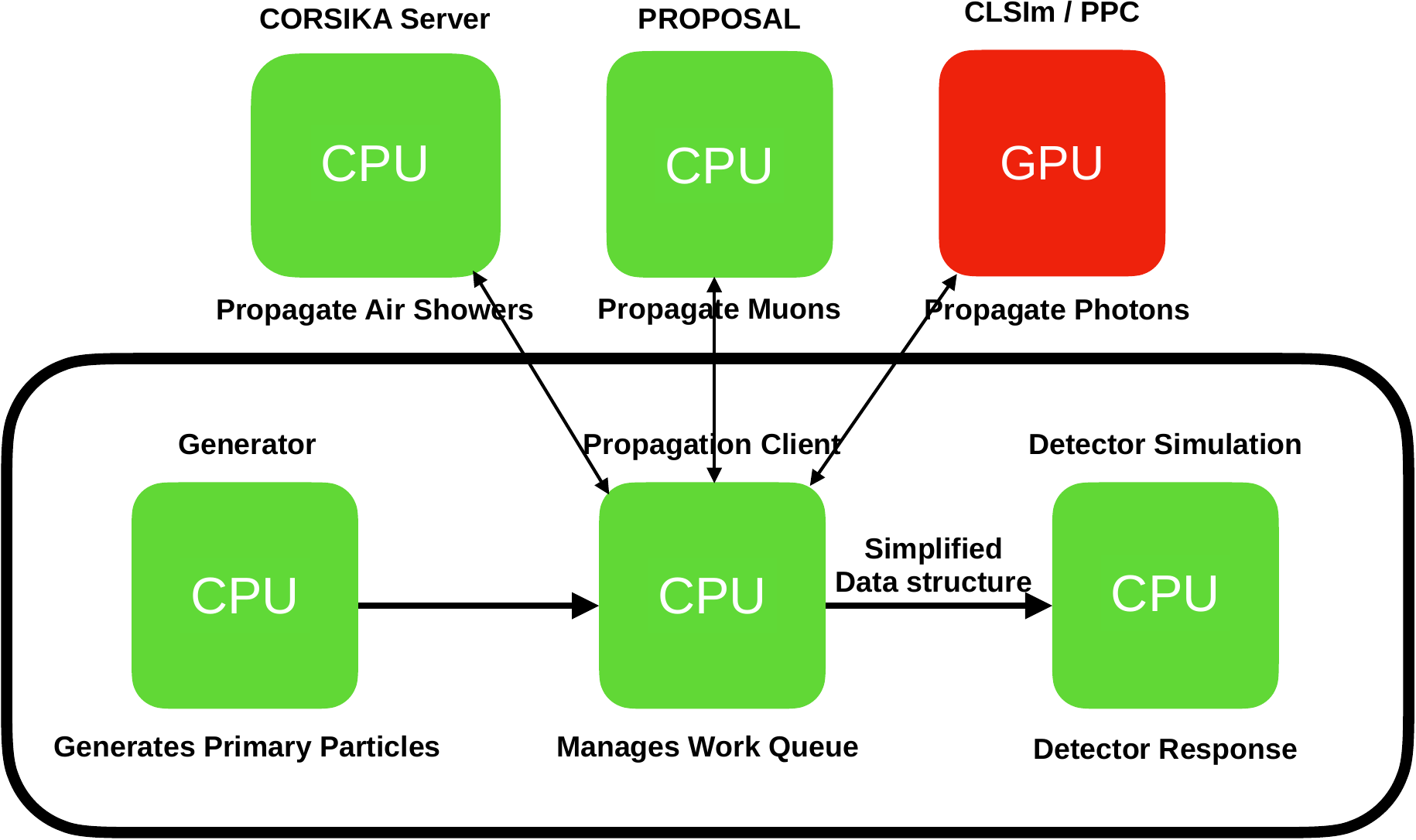}
  \caption{
    The new client--server simulation is shown in a simplified block diagram.
    Events are generated by a simple \textsc{IceTray} module.
    The event is then handled by the propagation client which uses a work queue to 
    store information from multiple events.
    Work is handed off as needed to servers which handle air shower, muon, and photon propagation. 
    After hits are generated they are handed off to the detector simulation.  
    }
  \label{block-new}
\end{figure}

\section{New Client--Server Simulation Chain}
\label{client-server}

With all these issues in mind, we undertook a significant refactor of our simulation chain.
This involves a new client--server model for air shower, muon, and photon propagation.
A block diagram is shown in Figure \ref{block-new}.
The propagation client now holds a queue of all particle propagation which needs to occur
and it connects to a number of servers using interprocess communications (IPC).
The first step of the chain is that the direction, location, and energy of the cosmic ray primaries 
are now sampled using a new \textsc{IceTray} module, rather than the distributions available from \textsc{Corsika}.

Since we do not rely on \textsc{Corsika} to generate primary particles, a new \textsc{IceTray} module was written.
This generator module simply produces primary particles on the surface of a cylinder slightly larger than the 
fiducial volume of IceCube. 

The heart of the new chain is a new \textsc{IceTray} module, referred to as the propagation client,
which stores all the particles to be tracked in a queue.
At each step of particle processing it is sent to the appropriate server based on the type of particle. 
The server returns the appropriate daughter particle in addition to any light created by the 
propagation of that particle.
The daughter particles are added to the queue and the photon emitting steps are stored in a separate queue.
When the step queue is sufficiently full, a block of steps is sent to the propagation server for processing as a batch.
The server aggregates steps from multiple events and clients in order to keep the GPU as busy as possible.

Since the queue does not need to store all of the intermediate particles at the same time,
there are significant memory savings. 
We also chose to take the opportunity presented by refactoring the code to replace
the previous data structure used to store hits with a more efficient one.
The old code used a data structure which stored a lot of diagnostic information
such as initial direction and number of scatters experienced.
This information is useful for debugging simulation code but is unnecessary for production.
It was replaced with a new object which only stores the ID of the particle
that produced it, the time of the hit, and the weight.
Together these two changes result in a significant memory savings.

The above changes are already improvements, but will not allow us to run propagation code on the same node
as the GPU unless we reduce the amount of CPU needed to the point where \textsc{Corsika}
and \textsc{Proposal} are running faster than CLSim.
To reduce the amount of CPU required we made two changes to how we process \textsc{Corsika}.
Both of these changes were only made possible by an extension to \textsc{Corsika} written by 
D. Baack~\cite{CORSIKAserverURL} which allows two-way communication with a running \textsc{Corsika} instance.

\begin{figure}[t]
  \centering     
  \includegraphics[width=0.49\textwidth,clip]{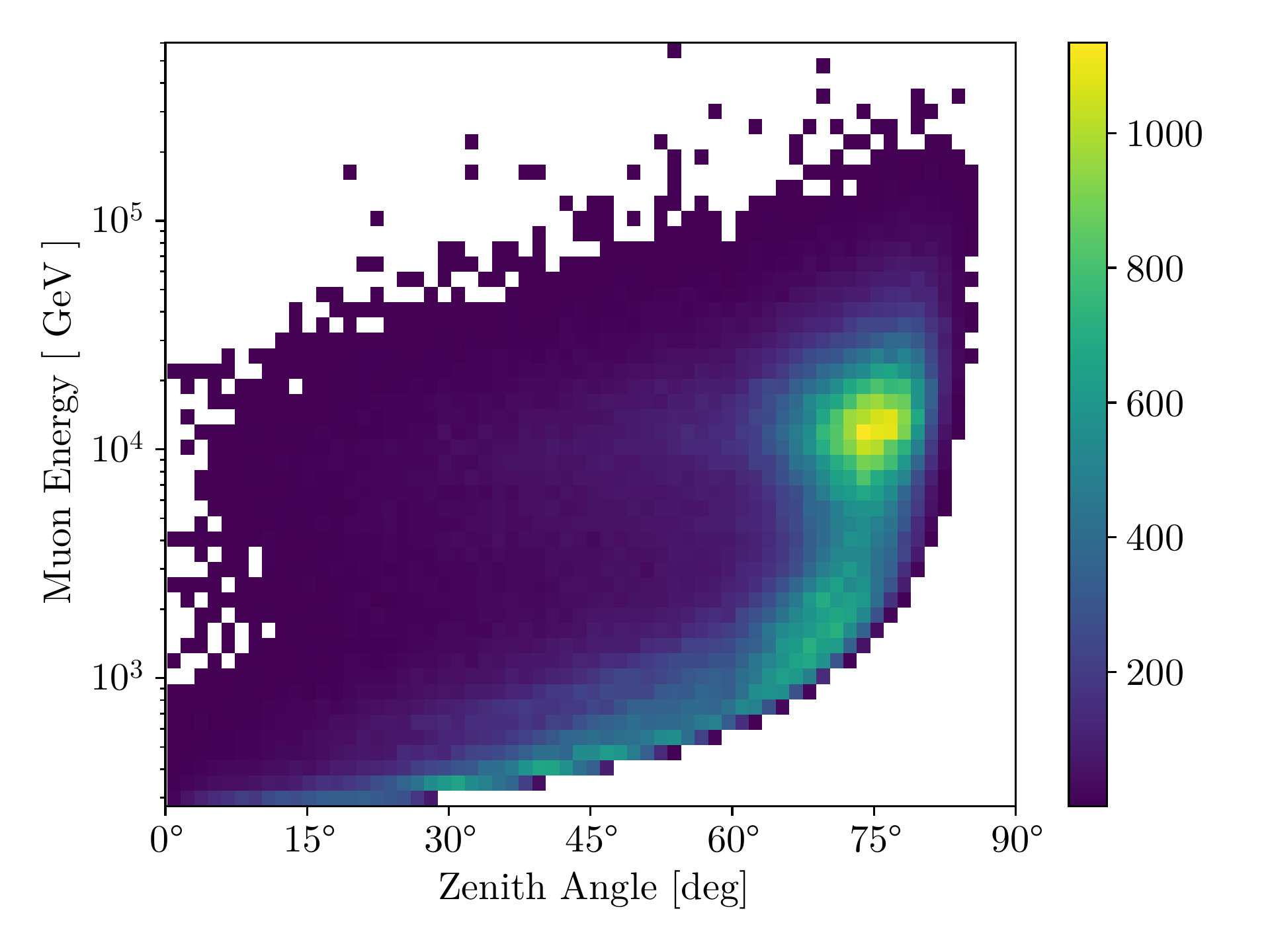}
  \includegraphics[width=0.49\textwidth,clip]{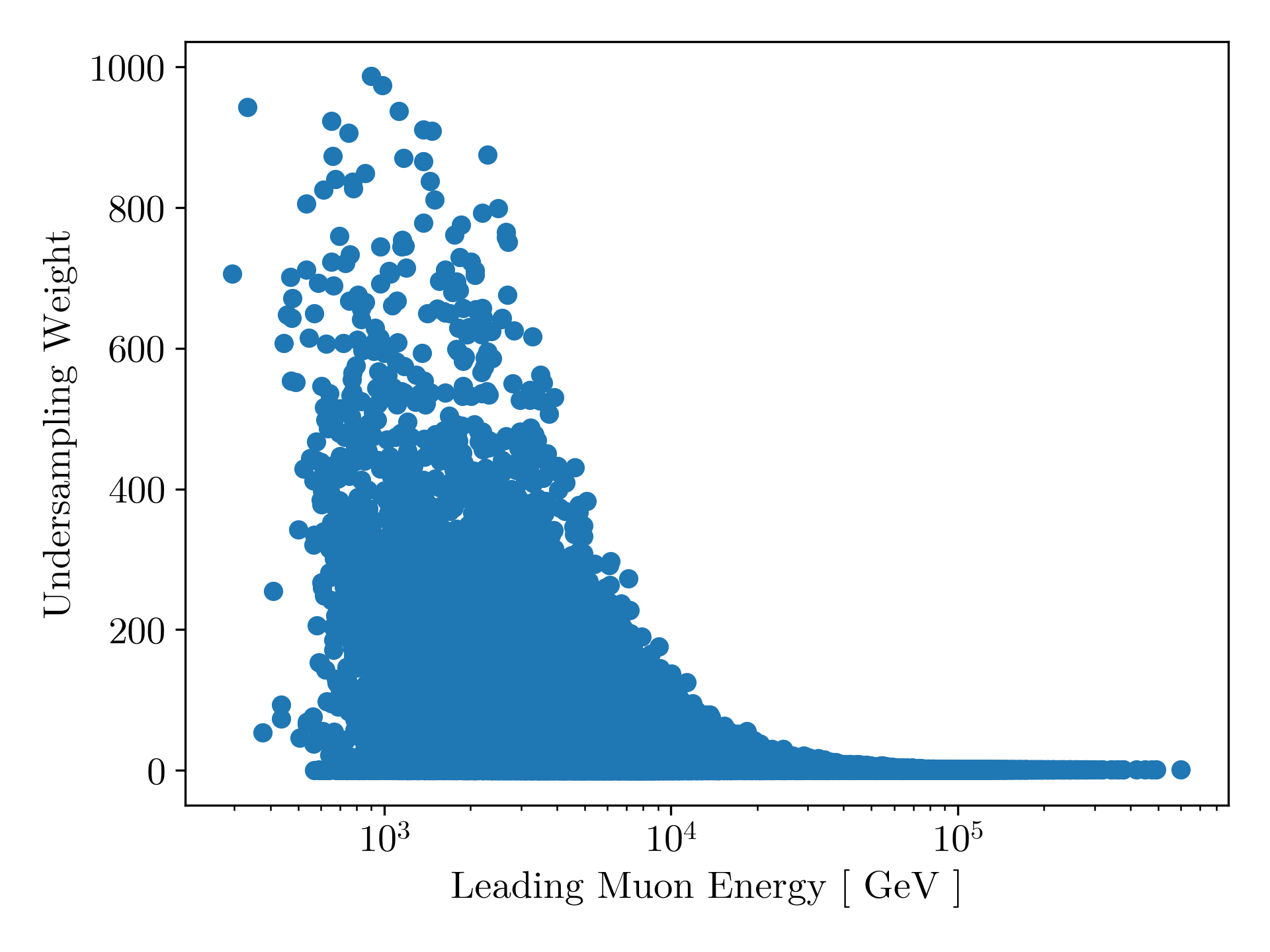}
  \caption{
    \textbf{Left:} The amount of events that generate light in the detector as a function
    of zenith angle and Muon Energy. Muons in the bottom right of the plot 
    do not have enough energy to make it to the detector volume and can be safely
    removed from the simulation without affecting the result.
    \textbf{Right:} The undersampling weight as a function of leading muon energy.
    Muons above 100\,TeV are not undersampled, whereas muons below 1\,TeV are undersampled 
    with a probability of 1 in 1000.
    Muons with intermediate energy are undersampled at an intermediate rate.
    }
  \label{leading_edge}
\end{figure}

The first speedup was to set the muon energy threshold based on the zenith angle of the shower.
CORSIKA takes the minimum energy before it halts propagation of muons as a parameter.
In traditional \textsc{Corsika}, the  parameter is fixed for all events,
which for extensive air shower arrays and other typical \textsc{Corsika} applications is rarely a problem.
This can, however, be a significant waste of resources for IceCube,
where the chance of a muon reaching the detector is highly dependent on slant depth as shown in Figure \ref{leading_edge}.
However, the IPC extension allows us to use a different configuration for every shower.
For highly inclined showers the overburden is significantly greater and the muon threshold
can be set much higher than what is needed for a vertical shower. 

The second speedup is related to the fact that most muons do not have sufficient
energy to reach the detector.
In a particular shower the muon most likely to reach the detector is the so-called leading muon,
the muon encountered by traversing the particle tree by always following the highest energy branch.
We created an undersampling scheme where we preferentially select 
showers with high energy leading muons and halt the calculation of the remaining showers.
The scheme has a single parameter, the undersampling factor $s$.
For each shower, this undersampling factor is used to find a threshold energy
where the probability of an air shower producing one or more muons of at least that energy is $1/s$,
using a parameterization of the average muon yield.
The shower is then simulated in descending order of energy while keeping track of the
highest-energy muon observed so far.
If it is above the critical energy, the simulation switches to depth-first propagation and runs to completion.
Otherwise, the simulation is halted according to the probability that the shower would
produce one or more muons of the current energy,
relative to the probability of producing one or more muons above the threshold energy.
For maximum muon energies far below the threshold energy, this probability converges to $1/s$.
For showers that run to completion, the survival probability is used to
weight the remaining showers up to account for those that were stopped.
This scheme allows us to efficiently sample signal-mimicking air showers while
retaining a subset of more typical showers to fill out the phase space.
A sample of weights is shown in the right side of Figure \ref{leading_edge}.
Finding the optimum value for the undersampling factor is discussed below.

\begin{figure}[t]
  \centering
  \includegraphics[width=0.49\textwidth,clip]{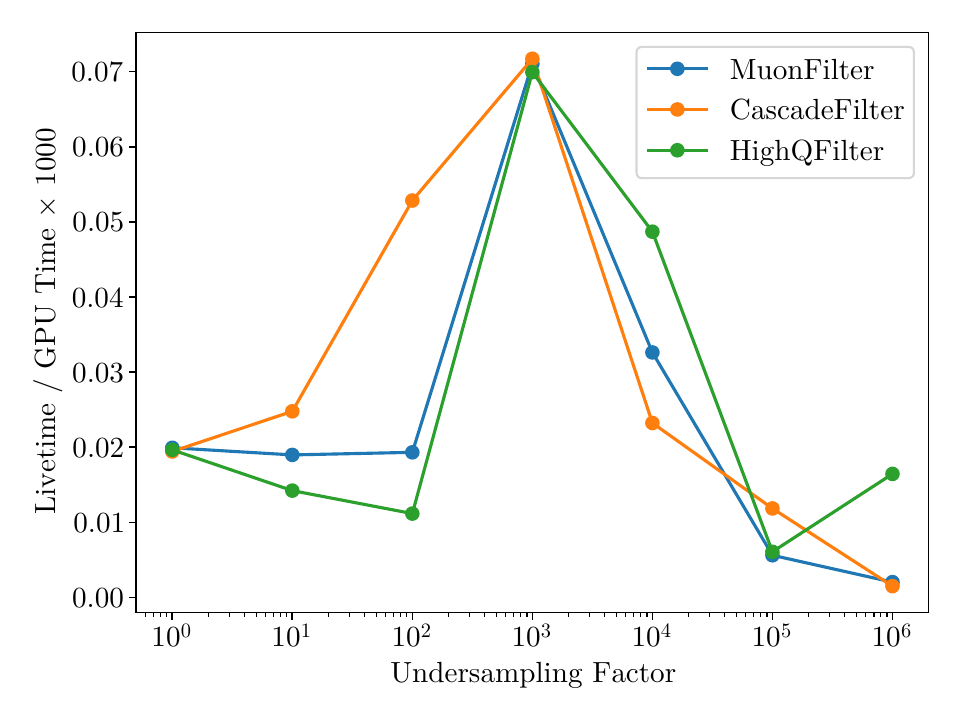}
  \includegraphics[width=0.49\textwidth,clip]{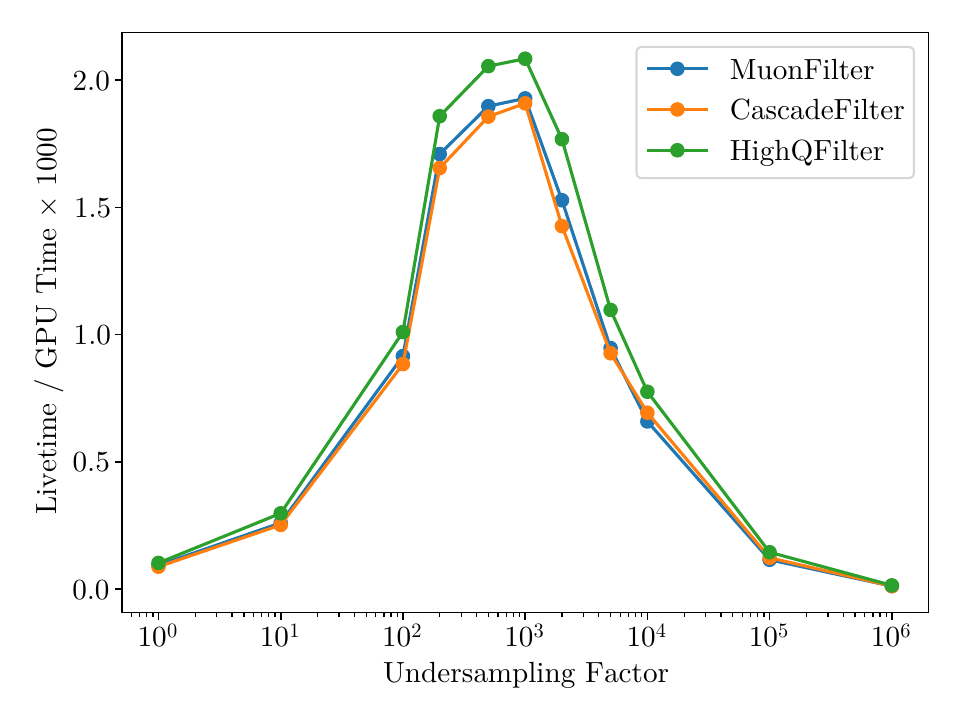}
  \caption{
    Tests performed to determine the optimum undersampling factor.
    \textsc{MuonFilter}, \textsc{CascadeFilter}, and \textsc{HighQFilter}
    are different selections of events based on computations performed at the South Pole
    that are intended to identify events likely to be neutrinos.
    A high factor will result in too few showers making it to 
    the photon propagation step, starving the GPU of photons.
    The left plot is using a primary particle range from 30\,TeV to 1\,PeV, 
    while the right plot is using 1\,PeV to 10\,EeV.
    Too low of a factor will result in most of the CPU being spent on showers
    which do not generate muons that emit light in the detector volume.
    The vertical axes show the amount of simulation \textsc{Livetime} divided by
    the amount of GPU time it takes to generate it; higher is better.
    A clear preference for a factor of $10^3$ is clearly seen in both energy ranges.
  }
  \label{undersampling_fig}
\end{figure}

\begin{table}[t!]
  \centering

  \caption{Summary of improvements between old simulation and new parallel simulation.
  \textsc{Completed Walltime} is defined as the Walltime of Completed Jobs
  divided by Walltime of Completed and Failed jobs.}
  \begin{tabular}{lcc}
                        & Old Chain    & New Chain \\
    \hline
    Time Per Shower    & $101 \mu{}s$ & $3.70 \mu{}s$ \\
    Memory/Core	       & 7GB          & 1GB           \\
    Completed Walltime & 60\%         & 84\%          \\
    GPU utilization    & 45\%         & 80\%          \\
  \hline
  \end{tabular}
  \label{comparison_table}
\end{table}

While the above changes significantly reduce the amount of computation needed to propagate
particles, a single CPU core will still struggle to produce
enough showers to fully utilize the GPU.
Since a typical cluster computing configuration has 4 or 8 CPU cores available for every GPU, 
we optimized our setup for these conditions.
Parallelization is achieved by running multiple instances of \textsc{IceTray}, 
each with their own generator and \textsc{Proposal}, but sharing a single
photon propagator server.
The more instances of \textsc{IceTray}, the easier it is to keep the GPU busy,
but the less likely to get an available slot in the cluster.
We tested parallelization by creating 1, 2, 4, 8, or 16 instances of \textsc{IceTray},
with 8 instances being optimal for our computing resources.

In addition to the memory savings described above, 
parallel \textsc{IceTray} also distributes the memory resources across the allocated cores
preventing exceeding memory limits.
Since the amount of memory allocated for an event is highly stochastic, 8 instances sharing
8\,GB of memory is significantly less likely to exceed the limit than 8 separate jobs
that allocate 1\,GB each.

\section{Results}
\label{results}

\begin{figure}[t]
  \centering
  \includegraphics[width=1\textwidth,clip]{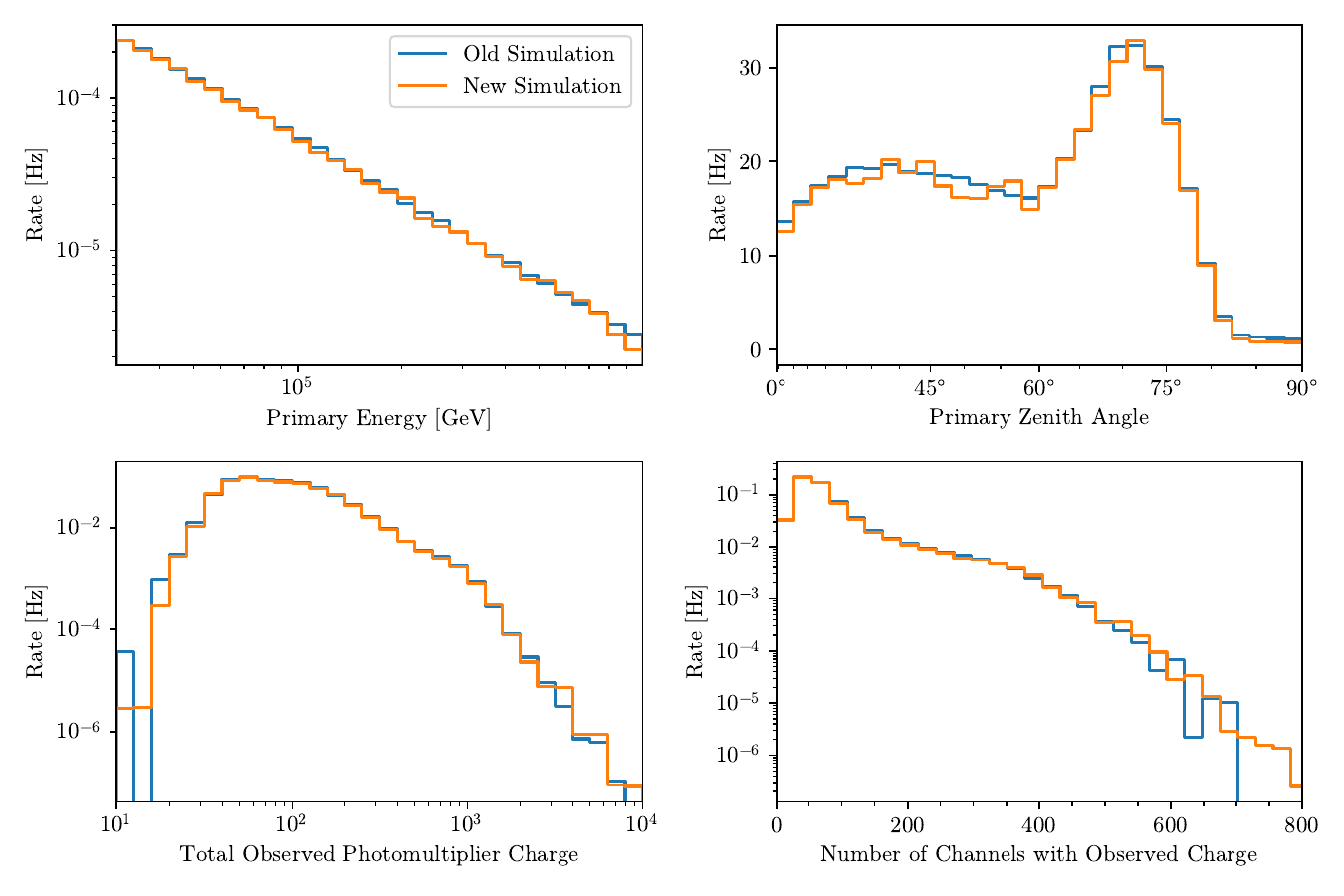}
  \caption{
    Comparison between the old simulation chain (shown in blue) versus the 
    new chain (shown in orange.)
    Very good agreement is seen, the discrepancy is believed to originate
    from different settings used for the atmospheric model used by \textsc{Corsika}
    for propagating cosmic ray air showers. 
  }
  \label{primary-energy}
\end{figure}

To optimize the value of the undersampling factor, we tested different values on Open Science Grid,
the results of which are shown in Figure \ref{undersampling_fig}.
A clear preference for a value of $10^3$ is seen,
which results in a $20\times$ performance improvement in producing \textsc{Livetime} compared to no undersampling.

To summarize, the client--server chain exhibits multiple benefits at all energy scales:
At low energies the Input/Output bottlenecks that were present are removed
by using IPC to communicate with \textsc{Corsika} instead of copying large output files between compute nodes.
At medium energies the calculations requiring CPUs are reduced to the point where
they can be performed on the same node and in parallel with the GPU calculations.
And at high energies the chances of a job halting memory spike is significantly reduced.
As seen in Figure \ref{primary-energy} very good agreement is seen in the observables of the 
new and old simulation.
A comparison of the performance between the old simulation chain and the new client--server chain
is shown in Table \ref{comparison_table}.

\section{Future Directions}
\label{future}
Going forward, the new simulation chain allows for a number of possible improvements 
that would not have been available with the old chain.

Since the simulation chain is now freed from relying on \textsc{Corsika} to generate 
primary directions, another way to increase the efficiency of simulation 
would be to oversample certain regions of interest.
Many IceCube analyses have identified preferred air shower trajectories
that have a higher chance of mimicking signal after quality cuts than others.

Additional oversampling can also be applied to the development of the air shower in \textsc{Corsika}.
For tau neutrino analyses and other high energy analyses, understanding the contribution
from prompt decays of particles like charmed hadrons in the air shower is important,
  so we could oversample such showers.
There also exist cosmic ray analyses which look for showers with high transverse momentum
interactions. To study these we could also oversample such showers.

Other techniques also become available with the new simulation chain 
which were not available with traditional \textsc{Corsika}.
It required that all of the showers in the same run be simulated with
the same configuration. 
This would require that, for running combined IceTop and InIce Simulation, 
the electromagnetic component of the shower be simulated for all events.
Separate configurations can now be sent based on 
the location of the shower, such that the electromagnetic component is only simulated
for showers which intersect IceTop.

Another technique that is now available would be to 
preferentially simulate air showers that produce high-energy neutrinos.
Such simulations would be useful for studies of downward-going atmospheric neutrinos,
which may be observed together with muons from the same shower.
Finally, the improved efficiency of the new simulation chain can be used
for studies which parameterize the muon flux at different detector depths. 
These are needed for \textsc{MuonGun}, a simulation generator which starts with the 
muon at detector depths, adapted from \textsc{MuPage}~\cite{MUPAGE}.
Further improvements are also expected with \textsc{Corsika~8}~\cite{CORSIKA8a,CORSIKA8b}
which has been rewritten in C++.
This rewrite will allow better incorporation of our client--server code than
the previous monolithic code of \textsc{Corsika}~7.

\end{document}